\documentclass{vgtc}                          % final (conference style)
\graphicspath{{figures/}{pictures/}{images/}{./}} % where to search for the images

\usepackage{times}                     % we use Times as the main font
\usepackage{tabu}                      % only used for the table example
\usepackage{booktabs}                  % only used for the table example
\usepackage{lipsum}                    % used to generate placeholder text
\usepackage{mwe}                       % used to generate placeholder figures

\usepackage{mathptmx}                  % use matching math font
\usepackage{graphicx}
\usepackage{adjustbox}
\usepackage[table,svgnames]{xcolor}
\usepackage{pifont}
\usepackage{dcolumn}
\usepackage{siunitx}
\usepackage{soul}
\usepackage{enumitem}
\onlineid{0}

\vgtccategory{Research}

\vgtcinsertpkg

\title{Generating Analytic Specifications for Data Visualization from Natural Language Queries using Large Language Models}

\author{%
    \begin{tabular}{ccc}
        Subham Sah$^\pi$\thanks{e-mail: ssah1@uncc.edu} & Rishab Mitra$^\pi$\thanks{e-mail: rmitra34@gatech.edu} & Arpit Narechania$^\pi$\thanks{e-mail: arpitnarechania@gatech.edu} \\
        \scriptsize UNC Charlotte & \scriptsize Georgia Institute of Technology & \scriptsize Georgia Institute of Technology \\
        \\[-8pt] % Adjust the vertical gap here
        Alex Endert\thanks{e-mail: endert@gatech.edu} & John Stasko\thanks{e-mail: stasko@cc.gatech.edu} & Wenwen Dou\thanks{e-mail: wdou1@uncc.edu\newline\indent$^\pi$authors contributed equally} \\
        \scriptsize Georgia Institute of Technology & \scriptsize Georgia Institute of Technology & \scriptsize UNC Charlotte \\
    \end{tabular}%
}

\abstract{
Recently, large language models (LLMs) have shown great promise in translating natural language (NL) queries into visualizations, but their ``black-box'' nature often limits explainability and debuggability.
In response, we present a comprehensive text prompt that, given a tabular dataset and an NL query about the dataset, generates an analytic specification including (detected) data attributes, (inferred) analytic tasks, and (recommended) visualizations.
This specification captures key aspects of the query translation process, affording both explainability and debuggability. 
For instance, it provides mappings from the detected entities to the corresponding phrases in the input query, as well as the specific visual design principles that determined the visualization recommendations.
Moreover, unlike prior LLM-based approaches, our prompt supports conversational interaction and ambiguity detection capabilities.
In this paper, we detail the iterative process of curating our prompt, present a preliminary performance evaluation using GPT-4, and discuss the strengths and limitations of LLMs at various stages of query translation. The prompt is open-source and integrated into NL4DV, a popular Python-based natural language toolkit for visualization, which can be accessed at \textbf{\url{https://nl4dv.github.io}}.
} % end of abstract


\begin{document}

\firstsection{Introduction and Background}

\maketitle

%% \section{Introduction} %for journal use above \firstsection{..} instead
Data visualization is an important component of data-driven storytelling~\cite{Knaflic.ch2}. 
However, existing tools for creating visualizations often require specialized knowledge, either through programming or using a graphical user interface (GUI), which limits authoring, customization, and analysis capabilities to experts.

One way to overcome this limitation and increase user access is by using natural language (NL) to create visualizations. Cox~et~al.~\cite{cox2001multi} first introduced the concept of creating visualizations from structured NL commands (NL2VIS). 
Since then, many natural language interfaces (NLIs) and toolkits for visualization have emerged that use keyword-based or semantic parsing-based approaches to interpret queries~\cite{gao2015datatone, yu2019flowsense, narechania2020nl4dv, mitra2022facilitating, valletto2018, inchorus2020, unspecifiedsetlur2019, Srinivasan_databreeze_2021, orko2018}.
NL4DV \cite{narechania2020nl4dv, mitra2022facilitating} is one such toolkit that utilizes a rules-based approach, employing dependency parsers like CoreNLP \cite{corenlp-manning-2014} to provide an analytic specification that includes detected attributes, inferred tasks, and visualization recommendations from an NL query and dataset. 
However, approaches like NL4DV require developers to create complex rules, which can limit the range and flexibility of input NL queries.
Advancements in natural language processing (NLP) and deep learning have further improved NL2VIS systems, which utilize transformers to interpret queries~\cite{ncnet,liu2021advisor}.

More recently, large language models (LLMs) like GPT-4~\cite{openai2024gpt4}, Claude~\cite{claude}, and Gemini~\cite{gemini} have been shown to effectively analyze and extract meaningful information, key concepts, relationships, and trends from unstructured textual data~\cite{minaee2024llmsurvey}. 
These capabilities have since been utilized for creative writing~\cite{gomez2023confederacy},
code generation~\cite{chen2021evaluating, llmsql}, dataset curation~\cite{ko2024natural}, and visualization creation~\cite{graphydeeplearning2022, poesia2022synchromesh, dibia2018data2vis, tian2023chartgpt}.
One notable LLM-based visualization system, chartGPT~\cite{tian2023chartgpt}, has outperformed a parsing-based system (NL4DV~\cite{narechania2020nl4dv}) and a deep-learning based system (ncNet~\cite{ncnet}).
In spite of their superior performance, LLM-based systems have certain documented limitations, such as providing insufficient explanations for the system's generated output~\cite{dibia2023lida} and being inconsistent in generating visualizations~\cite{chat2vis}. These unexplainable, uncertain systems impact transparency and trust, making it difficult for users to find and fix errors. In the NL to SQL domain, several explainable systems have already helped users identify and fix errors in the generated SQL queries~\cite{nl2sqlstudy, elgohary2021nl, narechania2021diy}, motivating this work for more explainable NL2VIS scenarios.

\begin{figure}[ht]% specify a combination of t, b, p, or h for top, bottom, on its own page, or here
  \centering % avoid the use of \begin{center}...\end{center} and use \centering instead (more compact)
  \includegraphics[width=\columnwidth, trim=2cm 2cm 3cm 2cm]{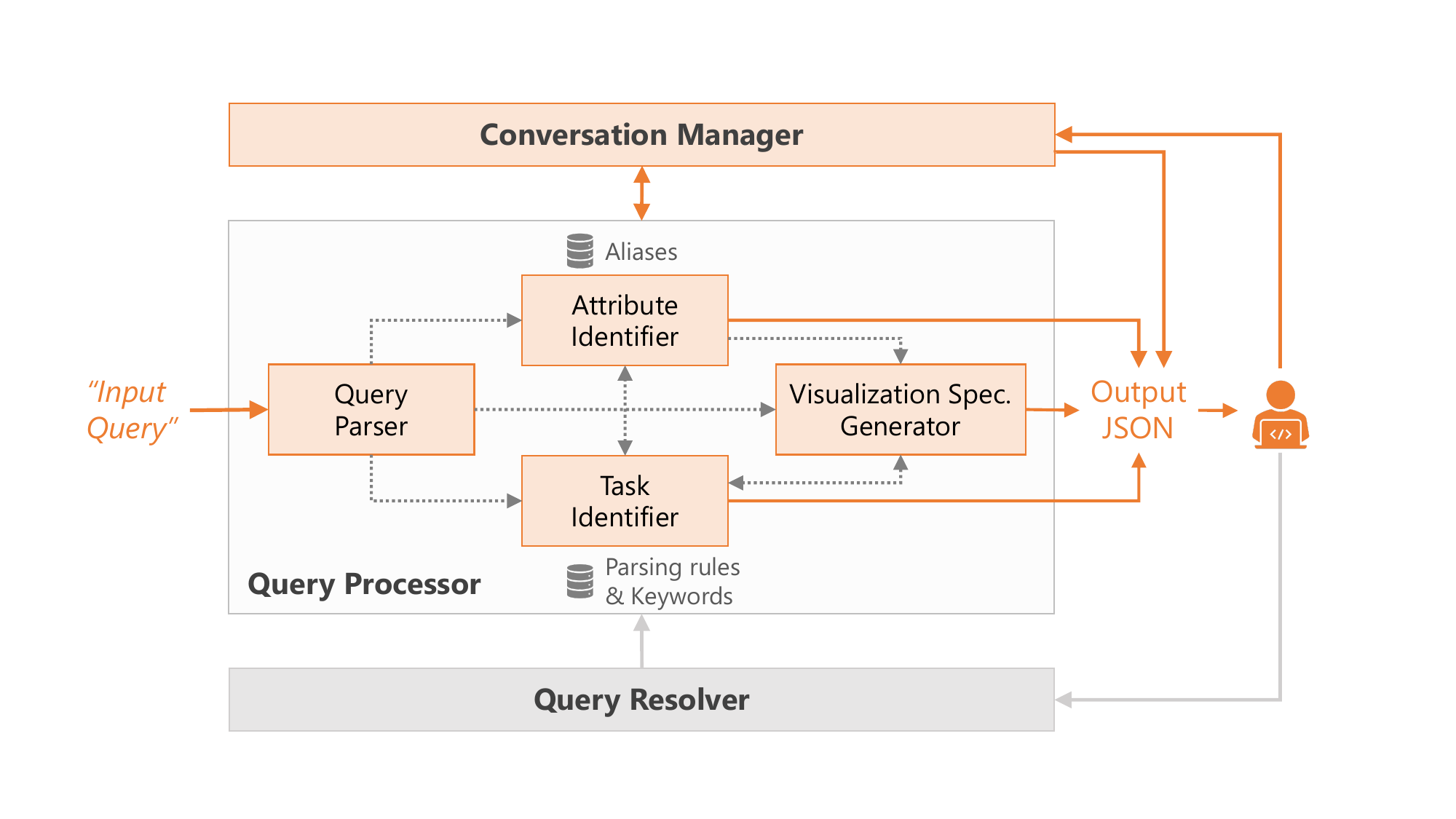}
  \caption{Architecture diagram of NL4DV~\cite{narechania2020nl4dv,mitra2022facilitating} highlighting the modules supported by NL4DV-LLM (in \textcolor{orange} {orange}), with arrows indicating the flow of information.
  }
  \label{fig:NL4DV-LLM architecture}
\end{figure}

In this work, we present a new LLM-based text prompt (NL4DV-LLM) that, like NL4DV~\cite{narechania2020nl4dv,mitra2022facilitating}, returns an analytic specification containing data attributes, analytic tasks, and relevant visualizations. 
This specification, presented as a structured JSON object (preferred by developers) or a step-by-step natural language explanation (preferred by users), affords explainability and debuggability by documenting key aspects of the query translation process. 
For example, it provides the mappings between the detected entities and the corresponding phrases in the input query as well as the specific visual design principles that determined the visualization recommendation. 
Furthermore, this prompt offers conversational interaction and ambiguity detection functionalities which are currently unsupported in other LLM-based NL2VIS systems. 
Essentially, users can ask follow-up queries to alter previously generated analytic specification(s) based on their evolving needs. 
If a query also contains ambiguities (e.g. a query phrase that can map to multiple data attributes), the prompt outputs multiple visualizations, one for each ambiguous entity.
However, the prompt does not support query resolution since it is a programmatic capability outside the scope of query translation~\cite{mitra2022facilitating}.
Figure \ref{fig:NL4DV-LLM architecture} highlights this key difference in the capabilities of NL4DV and NL4DV-LLM.

% \textcolor{red}{Why we sought to? We conduct...Reporting speech.}
We also conducted a preliminary evaluation of NL4DV-LLM against NL4DV using the NLVCorpus dataset~\cite{srinivasan2021nlvcorpus} and GPT-4~\cite{openai2024gpt4} as the prompt's engine. 
We found that in our corpus of 740 queries across three datasets, our prompt achieved an accuracy of 87.02\% compared to NL4DV's accuracy of 64.05\%.
However, on average, the LLM took around 25 seconds to generate analytic specifications, which can be a potentially unreasonable wait time for users. 
We discuss the tradeoffs and strengths of this prompt, which we hope guides future developments in natural language to visualization interfaces.
Our primary contributions are the following:
\begin{itemize}[nosep]
  \item An LLM-based NL2VIS text prompt (NL4DV-LLM) that translates NL queries about a tabular dataset into a comprehensive analytic specification that includes detected attributes, analytic tasks, and a recommended visualization.
  \item Description of our iterative process to curate the prompt. 
  \item Findings from a preliminary evaluation of NL4DV-LLM against NL4DV and a discussion on the prompt's strengths and tradeoffs.
\end{itemize}

\section{Developing the NL4DV-LLM Prompt}

To make LLM-based NL2VIS systems more explainable, we curated a text prompt, NL4DV-LLM. Given a tabular dataset and a natural language query about the dataset, this prompt produces a detailed analytic specification, including data attributes, analytic tasks, visualizations, and additional metadata explaining the translation process.
NL4DV-LLM's analytic specification essentially tries to replicate the output of NL4DV~\cite{narechania2020nl4dv}, a popular semantic parsing-based toolkit.
In this section, we describe the components of our prompt and the iterative process of engineering it.

\subsection{Prompt Components}

Figure \ref{fig:prompt} illustrates the various components of NL4DV-LLM along with several example visualization outputs on different kinds of input queries. We describe each component below.

\subsubsection{Analytic Tasks \& Visualization Design Knowledge}
NL4DV's output analytic specification includes Amar~et~al.'s~\cite{amar2005LowlevelCO} low-level components of analytic activity (i.e. ``analytic tasks'') as inferred from the input query.
To include these analytic tasks as part of NL4DV-LLM's output, we first probed GPT-4 to check if it has `learnt' the theory about these analytic tasks during its training.
Upon finding GPT-4 has \emph{not yet} learnt about analytic tasks, we decided to supply this knowledge in the prompt.
Specifically, we curated a structured JSON comprising the \textbf{Name}, \textbf{Description}, a \textbf{Pro Forma Abstract}, \textbf{Examples}, \textbf{Attribute Data Types and Visual Encodings}, \textbf{Attributes and Visual Encodings Description}, and \textbf{Recommended Visualization} for seven analytic tasks (\textit{Correlation, Distribution, Derived Value, Trend, Filter, Sort, and Find Extremum}).
The \textbf{Attribute Data Types and Visual Encodings} property details the preferred visual encodings (e.g., ``X axis'') and the datatypes (e.g., ``Quantitative'') that can be mapped to them; the \textbf{Attributes and Visual Encodings Description} provides instructions on how to encode these visual encodings and datatypes in Vega-Lite. The \textbf{Recommended Visualization} property specifies the visualization types most suitable for the given task, according to the design heuristics in NL4DV~\cite{narechania2020nl4dv}.

Using an ``in-context-learning''~\cite{li2024visualization} approach, we include this structured \textbf{Task JSON} as part of the main prompt.
We format this taxonomy as a JSON for succinctness and clarity, and to remove any potential ambiguities that may arise when utilizing NL.

\subsubsection{Conversational Interaction}
\begin{figure*}[p] % specify p for placing the figure on a separate page
  \centering 
  \includegraphics[width=\textwidth]{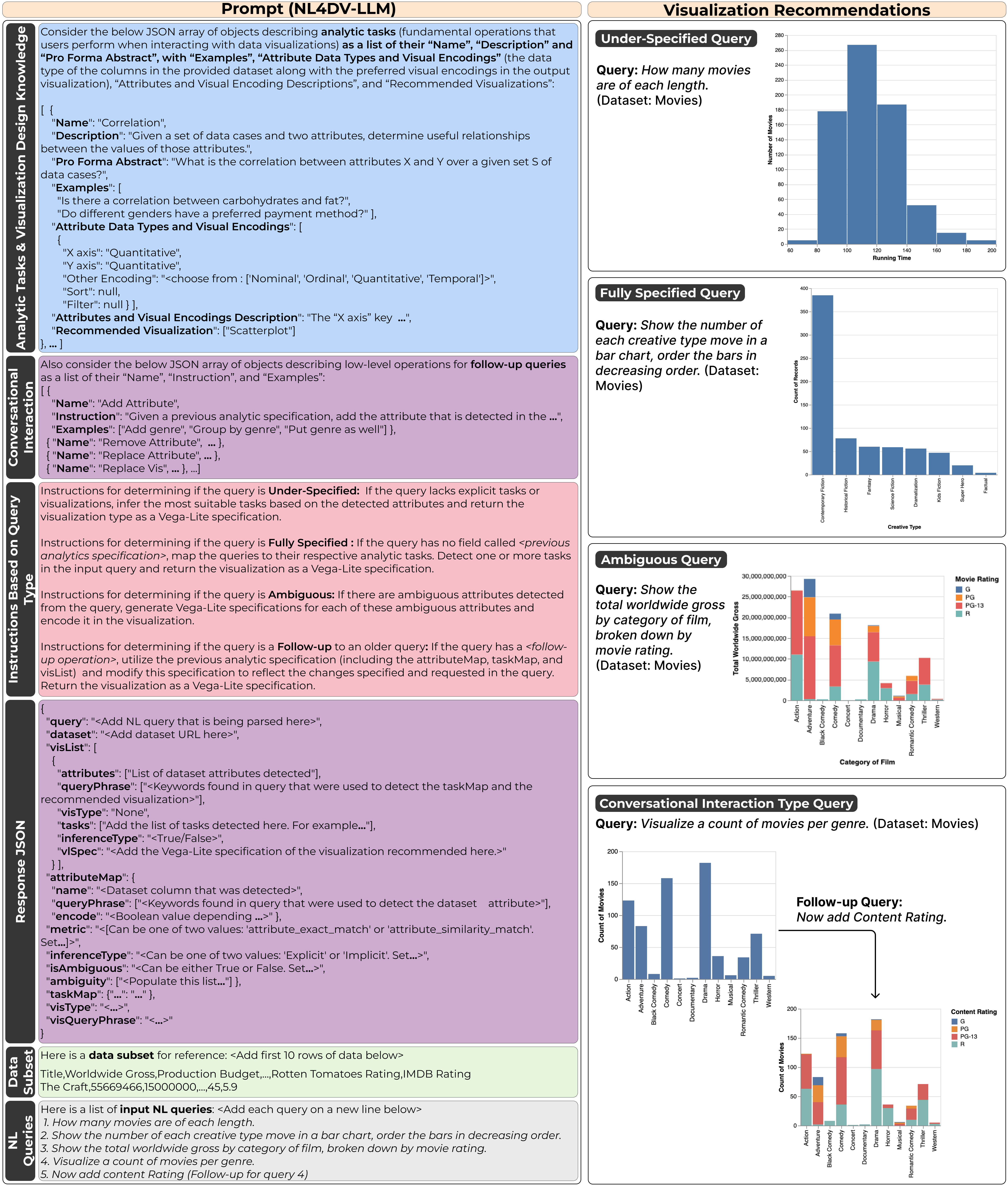}
  \caption{NL4DV-LLM prompt for Visualization Generation}
  \label{fig:prompt}
\end{figure*}

Similar to NL4DV~\cite{narechania2020nl4dv}, NL4DV-LLM supports modifications to previously generated visualizations via follow-up queries. NL4DV utilizes a unique follow-up taxonomy that classifies follow-up queries as one of three types: add, remove, replace for one or more components (specifically attributes, tasks, visualization types) of an analytic specification. We replicate this taxonomy through a JSON array, shown in Figure \ref{fig:prompt}, that contains its resultant permutations (e.g. add + analytic task), instructions describing the necessary steps to perform each operation, and follow-up query examples to provide context for each permutation.

\subsubsection{Instructions Based on Query Type}

The key instruction in our prompt to output an analytic specification is as follows:

\textit{``...classify the below natural language queries into the respective analytic tasks they map to. There can be one or more analytic tasks detected in the input natural language query. Return the visualization type in the form of a Vega-Lite specification where it reads data from the url above.''}

The instructions included in this component specify explicit steps on handling common NL query types: ambiguous queries, fully specified queries, underspecified queries, and follow-up queries. Figure \ref{fig:prompt} illustrates an example of each query type and its corresponding visualization.

\textbf{Underspecified Queries}: We utilize NL4DV's concept of underspecified queries, which is defined as queries that ``\textit{implicitly refer to tasks and visualizations}~\cite{narechania2020nl4dv}. If the query does not contain explicit references to tasks or visualizations, then the prompt instructs the LLM to utilize the design guidelines posited by the Analytic Task JSON, ``\textit{infer the task that is best suited with the detected attributes' datatypes}'', and ``\textit{generate a visualization specification using this inferred task and detected attributes}''.

\textbf{Fully Specified Queries}: We also define fully specified queries like NL4DV, where the NL query makes explicit references to at least one attribute, task, and visualization type~\cite{narechania2020nl4dv}. The key instruction described previously is sufficient to ensure coverage of fully specified queries; no other instructions are required.

\textbf{Ambiguous Queries:} The prompt defines ambiguous queries as queries ``\textit{with partial references to multiple data attributes.}'' In such cases, the prompt instructs outputting multiple visualizations, one for every attribute that a keyword potentially refers to, to maximally cover the user's intent.

\textbf{Follow-up Queries}: Since follow-up queries alter components (tasks, attributes, or visualization types) of a previously generated analytic specification rather than creating an entirely new one, the prompt includes another set of instructions to handle follow-up queries. For this query type, users must append a previously generated analytic specification to the end of the prompt. With this previously generated analytic specification, the primary instruction to handle follow-up queries is as follows:

\textit{``...classify the below natural language query into the respective follow-up operations they map to. Utilize the previous analytic specification (including the attributeMap, taskMap, and visList) and modify this specification to reflect the changes specified and requested in the natural language query. Return the visualization type in the form of a Vega-Lite specification where it reads data from the url above.''}

\subsubsection{Response JSON}
As shown in Figure \ref{fig:prompt},  NL4DV-LLM's output replicates NL4DV's, and contains an \textit{attributeMap} that is composed of the dataset attributes inferred from the natural language query, a \textit{taskMap} composed of the inferred analytic tasks, and a \textit{visList} that includes the Vega-Lite specifications relevant to the query~\cite{narechania2020nl4dv}. An example of this response is provided in the prompt itself, along with the instruction, ``\textit{Here is the JSON object that the response should be returned as}''. Also shown in Figure \ref{fig:prompt}, the example JSON object includes explicit instructions for each property, which constrains the output as much as possible, thereby increasing the prompt's consistency.

We provide an additional constraint in NL4DV-LLM's prompt by stating, \textit{``Do not include any additional prose in your response. I only want to see the JSON.''} This instruction ensures that the LLM only outputs the response JSON object, suitable for developers who are building NL2VIS applications on top of the prompt. Without this instruction, the LLM provides natural language explanations for its steps in formulating the response JSON for a given NL query. Users can opt to delete this sentence from the prompt if they would like to view additional explainable behavior from the LLM.

\subsubsection{Data Subset}

For the LLM to detect references to attributes and records in the input query, it needs access to the entire dataset.
However, due to a token limit imposed by the GPT-4 API, it is not possible to include the entire dataset as part of the prompt.
Consequently, we select all dataset headers (columns) and randomly subset ten records (rows) for the prompt.
While this choice alleviates the token-limit concern, it still has two limitations. First, for really wide datasets, the large number of columns may still exceed the token limit; and second, the ten sample rows may not be representative of the entire dataset and may result in false negatives in the query translating process.
To resolve cases where the amount of dataset columns exceeds the prompt's token limit, users can leverage methods in exploratory data analysis such as Linear Discriminant Analysis~\cite{lda2017} that only includes the most relevant columns as part of the prompt.

\subsubsection{NL Queries}
We finally conclude our prompt by including a list of one or more input natural language queries for processing.

\subsection{Prompt Iterations}
\begin{figure}[ht] % specify a combination of t, b, p, or h for top, bottom, on its own page, or here
  \centering % avoid the use of \begin{center}...\end{center} and use \centering instead (more compact)
  \includegraphics[width=\columnwidth]{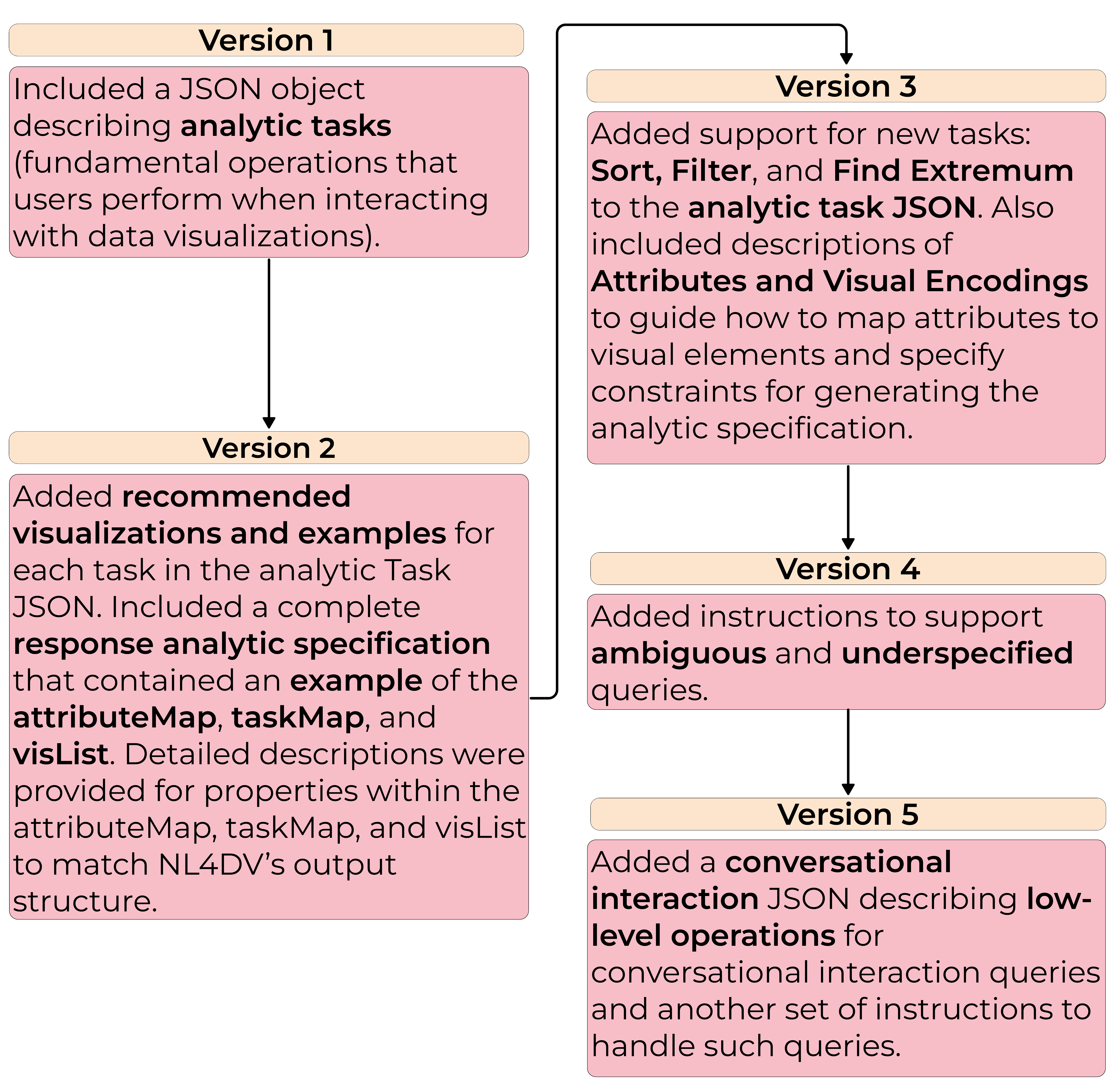}
  \caption{Illustration of our iterative prompt development process.}
  \label{fig:prompt version}
\end{figure}
% discuss how the current prompt evolved and why
Figure \ref{fig:prompt version} illustrates how our prompt 
% % evolved through 
underwent multiple iterations to improve its performance in analytic specification generation and support a wide variety of NL queries. The initial version of the prompt contained a JSON object describing a subset of analytic tasks found in Amar~et~al.~\cite{amar2005LowlevelCO}: \textit{Correlation}, \textit{Derived Value}, \textit{Distribution}, and \textit{Trend}. However, our prompt was often unable to detect the correct tasks for a given NL query. In addition, the prompt only outputted a Vega-Lite specification with no explanations whatsoever, affording limited explainability to the user. 
In Version 2, we augmented each task in the analytics task JSON object with example queries and recommended visualizations. 
In addition, we also included a sample analytic specification, similar to NL4DV's~\cite{narechania2020nl4dv}, including an \textit{attributeMap} and a \textit{taskMap}, to enhance the prompt's explainability to the user.
For example, users could now view the mappings between the detected dataset attributes and the corresponding phrase in the input NL query.

Next, in Version 3, we enhanced the analytic task JSON by introducing support for other analytic tasks, namely \textit{Sort, Filter, and Find Extremum}. 
An \textit{Attributes and Visual Encodings Description} property was also included for each task, providing the LLM heuristics on task inference for underspecified queries. 
Version 4 provided additional instructions to handle ambiguous and underspecified queries since earlier prompt versions were unable to generate accurate visualizations for such query types. 

Finally, in version 5, we introduced a conversational interaction JSON describing the low-level taxonomy introduced in Mitra~et~al.~\cite{mitra2022facilitating} and a distinct set of instructions to process follow-up queries. With this version, we ensure coverage of all natural language query types supported by NL4DV \cite{narechania2020nl4dv}.

\section{Preliminary Evaluation}
\subsection{Setup and Design}
To study our prompt's performance across different NL2VIS scenarios, we conducted a preliminary evaluation using GPT-4 as our prompt engine and NLVCorpus~\cite{srinivasan2021nlvcorpus} as our dataset benchmark (query corpus).
We chose GPT-4 as it was considered as the state-of-the-art LLM during the time of evaluation (February - March 2024).
We chose NLVCorpus for its human-generated utterance sets, which provide a realistic and robust representation of NL queries across three dataset domains: movies, cars, and superstore.

To evaluate our prompt's conversational interaction capabilities, we included follow-up queries from the NL4DV website~\cite{nl4dv}. We did not use NLVCorpus' sequential queries, as many requested unsupported aesthetic changes (e.g., \textit{Use major gridlines}). Our composite dataset comprised 740 queries. We executed all 740 queries via our prompt on GPT-4 and evaluated the outputs for correctness.
We defined a query to be correct if the generated analytic specification accurately captured the query intent and the resultant visualization included everything asked for in the query.

We provide the dataset corpus (used for evaluation), text prompt (NL4DV-LLM), annotations on the prompt's outputs, and a gallery of sample visualization outputs  which can be accessed at \textbf{\url{https://github.com/nl4dv/NL4DV-LLM-supplemental-material}}.

\subsection{Data Annotation Procedure}
We process all the queries in our evaluation corpus through NL4DV and NL4DV-LLM and record the response times for each query. 
We assess whether the output for each query should be deemed as ``accurate,'' signifying if the visualization matched the query's intent and included all analytic tasks and attributes requested by the query. If the output failed to meet these requirements, we marked the visualization as ``inaccurate'' and specified factors that caused the visualization's inaccuracy. Common reasons for inaccuracies were that the visualization was missing an analytic task, attribute, or encoded incorrect attribute(s). Furthermore, the systems sometimes did not generate outputs for certain queries. The outputs, in these cases, were marked as wholly inaccurate.

The first two co-authors served as annotators to evaluate the accuracy of the resulting visualizations. They first analyzed small subsets of the dataset to ensure that their analyses were fully calibrated with each other, creating a consistent standard in their analysis.  
Then they went through the systems' responses for the entire corpus of queries and determined if the responses precisely answered the queries. 
They compared their annotations with each other and discussed any discrepancies in their results to come to a consensus. In cases where the first two co-authors remained split in their decision, the third co-author would use their decision as the tiebreaker for the analysis. 

Since many of the input queries in the evaluation dataset could be considered as underspecified or ambiguous, multiple visualizations can be regarded as ``accurate'' for these types of queries. For example, in the movies' dataset, there can be multiple viable visualizations for the query ``\textit{Correlate budget, gross, and rating}'', where the keyword ``\textit{rating}'' can refer to the attributes ``\textit{Content Rating}'',  ``\textit{Rotten Tomatoes Rating}'', or ``\textit{IMDb Rating}''. 
For such cases, the annotators assessed if the output contained any valid interpretation of the query.
Therefore, the annotators opted not to refer to the ground truth provided for each query in NLVCorpus for the majority of their annotations, and instead only referred to the ground truth for any discrepancies in their results.

\subsection{Results and Discussion}
By manually analyzing and annotating every visualization generated by each NL4DV-LLM \& NL4DV, we discover that NL4DV-LLM with GPT-4 outperforms the NL4DV in accuracy but has significantly longer response times.

Out of the 740 queries, NL4DV-LLM generated 644 accurate responses, resulting in an accuracy rate of 87.02\%. NL4DV generated 474 accurate responses, resulting in an accuracy rate of 64.05\%.
The prompt's accuracy was mostly consistent across the three test datasets: 84.98\% on the cars' dataset, 89.89\% accuracy on the movies' dataset, and 86.5\% on the superstore dataset. 
However, NL4DV's accuracies on the movies' dataset (75.49\%) and cars dataset (70.73\%) were significantly higher than the superstore dataset (40\%), potentially due to the superstore dataset's complexity and high syntactic similarity among its attributes.
A previous evaluation reported that learning-based NL2VIS system ncNet~\cite{ncnet} had an accuracy of about 45\% and another LLM-based system chartGPT \cite{tian2023chartgpt} had an accuracy of about 79\%,  albeit on the nvBench dataset and a subtle distinction in its definition of accuracy \cite{tian2023chartgpt}. 
These findings suggest that our prompt's accuracy is comparable to, if not slightly higher than, previous systems.

Notably, our prompt handled a wide range of query structures and forms, including underspecified and ambiguous queries.
For example, our prompt was able to apply transformations and computations to dataset values to create new ``derived'' attributes (e.g. creating a new attribute \textit{Profit} from \textit{Production Budget} and \textit{Worldwide Gross} in the movies dataset), which semantic parsing-based approaches cannot generally support.
However, there were a few queries that resulted in inaccurate outputs.
Common reasons for inaccuracy were malformed response JSON objects or well-formed JSON objects with incorrect Vega-Lite syntax.
Such occurrences can undermine user trust, highlighting the need for future systems to implement appropriate safeguards.
In addition, certain outputs were misleading due to incorrect associations between the data attributes and the visual encodings. 
For example, for the query, ``\textit{Show total profit across genres,''} the y-axis might be labeled ``Total profit'' but actually use ``Worldwide Gross'' as the data field, making the visualization confusing to the user.

Lastly, NL4DV-LLM's average response time across all 740 queries was 25 seconds, significantly longer than NL4DV's average response time of 3 seconds~\cite{narechania2020nl4dv}. 
Such a long wait time can impact the prompt's usability. 
We attribute this high response time to our prompt's extensive size.
However, initial testing with another LLM (GPT-4o mini) demonstrated improved response times.

\section{Limitations and Future Work}

We aimed to make our prompt as explainable as possible by modeling the query interpretation process into the analytic specification JSON. This approach helped us explain system behavior across different datasets and queries. 
However, in the prompt, we also looked to incorporate confidence scores for the detected attributes, tasks, and visualization types in the response JSON, much like the confidence scores in NL4DV~\cite{narechania2020nl4dv}.
The confidence scores in NL4DV measure the semantic and syntactic similarity between the detected entities and the query phrases that they map to through functions like cosine similarity and Wu-Palmer scoring~\cite{wupalmer}.
However, whilst implementing this feature, we found that GPT-4 was unable to properly calculate token similarity scores or produce confidence scores meaningful to the user.
This shows a limitation in our prompt's explainability absent in parsing-based NL2VIS systems.

Next, we conducted our preliminary evaluation during February 2024 and March 2024. At the time, OpenAI's GPT-4 was considered as the state-of-the-art LLM. Since then, there have been a number of other LLMs released, like Claude-3.5~\cite{anthropic-sonnet} and GPT-4o~\cite{openai-gpt-4o-contributions} that have reported outperforming GPT-4 in a number of benchmarking tasks. Furthermore, GPT-4 itself may have undergone updates and changes, potentially affecting consistency, output accuracy, and response time. Future work is planned to conduct evaluations across different LLMs.

Finally, we conducted our preliminary evaluation using the NLVCorpus dataset, which only contains queries for three datasets. For a more thorough evaluation, we plan to utilize more diverse and comprehensive dataset benchmarks such as nvBench~\cite{luo2021natural}, which contains over 25,000 queries for 105 datasets.

\section{Conclusion}
We presented NL4DV-LLM, an LLM-based prompt that generates analytic specifications from a dataset subset and a natural language query, supporting traditional NL2VIS capabilities. Preliminary evaluation shows promise but also highlights limitations affecting trust and usability. We share our prompt curation experiences to guide future developers in NL2VIS tasks.

\acknowledgments{This material is based upon work supported by NSF CNS-2323795. We thank members of the Georgia Tech Visualization Lab and the Charlotte Visualization Center for their helpful feedback at different stages of this work. We used Google Scholar~\cite{googlescholar}, vitaLITy~\cite{vitality2022narechania}, and vitaLITy 2~\cite{an2024vitality2} to assist us during our literature review.}

\bibliographystyle{abbrv-doi}

\bibliography{template}
\end{document}